# Scalable Energy Efficient Location Aware Multicast Protocol for MANET (SEELAMP)

Pariza Kamboj, A.K.Sharma

**Abstract**—Multicast plays an important role in implementing the group communications in bandwidth scarce multihop mobile ad hoc networks. However, due to the dynamic topology of MANETs it is very difficult to build optimal multicast trees and maintaining group membership, making even more challenging to implement scalable and robust multicast in Mobile Ad hoc Networks (MANET). A scalable and energy efficient location aware multicast algorithm, called SEELAMP, for mobile ad hoc networks is presented in the paper that is based on creation of shared tree using the physical location of the nodes for the multicast sessions. It constructs a shared bi-directional multicast tree for its routing operations rather than a mesh, which helps in achieving more efficient multicast delivery. The algorithm uses the concept of small overlapped zones around each node for proactive topology maintenance with in the zone. Protocol depends on the location information obtained using a distributed location service, which effectively reduces the overheads for route searching and shared multicast tree maintenance. In this paper a new technique of local connectivity management is being proposed that attempts to improve the performance and reliability. It employs a preventive route reconfiguration to avoid the latency in case of link breakages and to prevent the network from splitting.

—————————— ◆ ——————————

## 1 INTRODUCTION

AN Ad hoc network consists of a collection of mobile hosts forming a dynamic multi-hop autonomous network [3] without the intervention of any centralized access point or fixed infrastructure. Multicast has great impact in mobile networks because of their inherent broadcast capability. Using multicast instead of sending through multiple unicasts not only minimizes link consumption, but also reduces sender and router processing, communication costs and delivery delay [7].

Group communication is important in Mobile Ad Hoc Networks (MANET). Many ad hoc network applications which require close association of the member nodes depends on group communication.

Action directions given to the soldiers in a battlefield and communications required during a rescue operation are some examples of these applications. In addition, many routing protocols for wireless MANETs need a broadcast/multicast as a communication primitive to update their states and maintain the routes between nodes [18].

Multicast protocols can be categorized in tree based and mesh based protocols. Multicast tree structures are frail therefore need to be readjusted and repaired continuously as the connectivity changes. Even in wired networks building optimal multicast trees and maintaining group membership information is challenging. Bandwidth scarcity, limited power resource, and above all dynamicity of topology in a mobile ad hoc network make the multicast protocol design predominantly challenging than that for wired network.

The proposed protocol, a scalable and robust location aware multicast algorithm, called SEELAMP uses the concept of zone and constructs a shared bi-directional multicast tree for its routing operations rather than a mesh, which helps in achieving more efficient multicast delivery. Zone building, multicast tree construction and multicast packet forwarding depends on the location information obtained using a distributed location service, which effectively reduces the overheads for route searching and shared multicast tree maintenance [17]. In this paper a new technique of local connectivity management is being proposed that attempts to improve the performance and reliability. It employs a preventive route reconfiguration to avoid the latency in case of link breakages and to prevent the network from splitting.

The rest of the paper is organized as follows: Section 2 takes a look at the tree based multicast protocols classification for MANET and also emphasizes the problems lie in the existing multicast routing protocols. The proposed Scalable Energy Efficient Location Aware Multicast Protocol is discussed in Section 3. Section 4 analyses the performance of SEELAMP in comparison with other tree-based multicast protocols. Finally, section 5 summarizes the study of the work in conclusions.

————————————————

- *Pariza Kamboj is with the Deptt. of Computer Sc. Engineering Institute of Technology Management (ITM), Gurgaon – 122017, India.*
- *A.K. Sharma is with the Deptt. of Computer Engg., YMCA Institute of Engineering, Faridabad-121006, India.*

## 2 MULTICAST PROTOCOLS FOR MANETs

Several multicast protocols have been proposed for mobile ad hoc networks. Based on the network structure along which multicast packets are delivered to multiple receivers, multicast protocols can be broadly categorized into two types, namely tree-based multicast and mesh-based multicast. A tree based multicast routing protocol is either a source-tree or a shared-tree protocol. In a source tree based multicast routing protocol data packets are delivered through multiple source-based routing trees routed at the sources of the multicast session and in shared tree protocol data packets are delivered along a shared multicast tree for the whole multicast group. If the nodes in the network are highly dynamic, a large number of source trees might need reconstruction, causing excessive overhead in case of source-tree multicast [20]. On the other hand, shared tree multicast has lower control overhead because it needs to maintain only a single shared tree for all multicast sources and therefore is more scalable [1].

The tree structure is known for its efficiency in utilizing the network resource optimally, while tree based protocols are generally more efficient in terms of data transmission. Mesh based protocols are more robust against topology changes due to availability of many redundant paths between mobile nodes and result in high packet delivery ratio. On the other hand, multicast mesh does not perform well in terms of energy efficiency because mesh-based protocols depend on broadcast flooding within the mesh and therefore, involving many more forwarding nodes than multicast trees. In summary, the broadcast forwarding in mesh based protocols produces redundant links, which improves the packet delivery ratio but spends more energy than the tree-based multicast. The shared tree approach has some other drawbacks. The paths are non-optimal and traffic is concentrated on the shared tree, rather than being evenly distributed across the network. They are not robust to mobility as there is no back up path between a source and a destination, besides that, all shared tree based protocols need a group leader (or a core or a rendezvous point) to maintain group information and to create multicast trees. A multicast packet is delivered to all the receivers belong to a group along a network structure such as tree or mesh, which is constructed once a multicast group is formed. However, due to node mobility the network structure is fragile and thus, the multicast packet may not be delivered to some members. To compensate this problem and to improve the packet delivery ratio, multicast protocols for MANETs usually employ control packets to periodically refresh the network structure [14].

### 2.1 Comparison of Multicast Protocols

Ad hoc Multicast Routing (AMRoute) [9] and Lightweight Adaptive Multicast (LAM) [19] are tree based protocols, in which a shared tree is constructed for the delivery of multicast packets to the entire multicast group. In AMRoute protocol a bi-directional shared multicast tree is created involving only the group members. The tree links are created as unicast tunnels between the tree members. The problem with AMRoute is that it depends heavily on an underlying unicast protocol for creating these unicast tunnels. The LAM protocol depends on Temporally-Ordered Routing Algorithm (TORA) for route finding ability and cannot operate independently. An advantage of LAM is that, it reduces the amount of control overhead generated for route finding, due to its tight coupling with TORA. However, due to shared tree structure these protocols have the disadvantage of their dependency on a core node, thus have a central point of failure. CAMP [2] and On-Demand Multicast Routing Protocol (ODMRP) [21] are well-known examples of mesh-based multicast routing protocols. They enhance the robustness by providing redundant paths between the source and destination pairs. The mesh is created at the cost of higher forwarding overhead. CAMP illustrates a proactive mesh based protocol. On the other hand, in ODMRP, the mesh is created using the forwarding group concept and a reactive approach is followed to keep the forwarding group current [20]. The main disadvantage with mesh based protocols is the excessive overhead incurred in keeping the forwarding group current and in the global flooding of the JOINREQUEST packets. Every multicast routing protocol is having some or the other problem, hence suitable to specific kind of environment. To overcome from these problems and to make an environment independent protocol, a hybrid approach is needed that can control the power resource, reduce the overhead and also enhance the scalability at the same time. Moreover, the location advantage of the nodes can further improve the performance of the protocol manifolds. Based on this view we have designed a new multicast routing protocol named Scalable Energy Efficient Location Aware Multicast Protocol (SEELAMP) with reduced overhead.

## 3 PROPOSED PROTOCOL SEELAMP

This section introduces a new multicast protocol, Scalable Energy Efficient Location Aware Multicast Protocol (SEELAMP), which follows the hybrid approach, a mix of proactive and reactive routing using the physical location of the nodes. Like some other multicast protocols neither it depends on any underlying unicast protocol nor on a core node (root node) as the backup root node provides support in case of primary root node failure. The protocol reduces the total energy consumption as well as improves the performance than a conventional shared tree based protocol.

### 3.1 Backup Root Shared Tree based Multicast

In case of shared multicast tree the protocol depends on a core or root node to maintain the group information which makes the root node easily overloaded. Due to this shared tree multicast is particularly not suitable from energy balancing point of view because the root of the tree takes on more responsibility for routing, consumes more battery energy, and stops working earlier than other nodes. This leads to MANET partitioning as well as reduced network lifetime [22].

In case of shared tree multicast protocol if the primary root node goes off then the whole multicast tree is disconnected into a number of partitions which consumes a lot of wireless bandwidth for searching the group leaders of all the partitions. To alleviate this problem, SEELAMP uses a shared multicast tree with backup root node instead of a single root node. Creation of a backup root node enhances the performance of the multicast tree and also lessens the load on the primary root node. In case of primary root node failure the backup root node takes over, therefore, reduces the dependency on a single root node. This facilitates a great reduction in tree maintenance overheads and tree re-construction. Selection of backup root node is done from the neighbor nodes of the primary root node on the basis of stability and battery status. A node with slow movement and more power status is chosen to be the backup root node. Selecting a suitable node as backup root node not only serves the purpose of standby root node but also defer the early possibility of searching the backup root node again in case of power failure or movement of the existing backup root node.

The proposed scheme is explained with the help of an example shown in Figure 1. In the given tree R is the primary root node. The root node searches the nearest node by comparing the nodes' locations in its neighborhood with lowest speed and good power status and found a node B suitable for the purpose of backup root node. If the root node does not found any neighbor node with the required criterion then the selection process is delayed by some random time and after that the backup root node search process starts again. The selection process may lead to slight delay but improves overall efficiency of the protocol by selecting a suitable node as backup node.

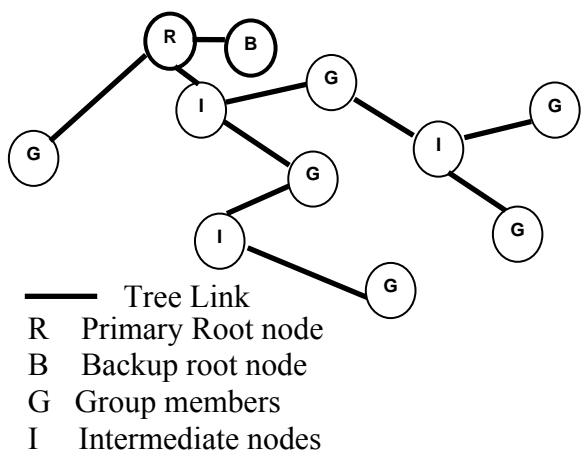

―――   Tree Link
R   Primary Root node
B   Backup root node
G   Group members
I   Intermediate nodes

Fig :1 Shared Multicast Tree with Backup Root

Figure 1 shows an example of a multicast tree. The tree consists of a root node (R), Backup root node (B), three intermediate nodes (I), six member nodes of a multicast group, and nine tree links. A multicast packet is delivered from the root node R to all the six group members. Using the zone routing every tree member unicasts the multicast packet only to the neighbor tree members, thus saves a lot many transmissions otherwise required in case of broadcasts.

### 3.2 Location Information of nodes

The routing performance can be significantly improved by utilizing location information of nodes in communication e.g., if a sender node knows the location of the tree member, it can find out the route to the tree member using diffused routing by forwarding the packet in the relative direction in hopes of getting it there quickly therefore communication delay can be minimized with location information. A node can use Global Positioning System (GPS) to obtain its geographic location information. The locations of other nodes can be obtained by implementing a distributed location service [6]. However, in practice, it is difficult to find/maintain node locations with accuracy in an ad hoc environment where nodes move around. Some well-known location-based routing algorithms are location-aided routing (LAR) protocol [29], distance routing effect algorithm for mobility (DREAM) [15], and grid location system (GLS) [12]. In SEELAMP a location service algorithm using swarm intelligence techniques, called SLS (Swarm location service) [11] developed by Bluetronix is used to provide location information to all mobile ad hoc nodes and the geographical information obtained is used to limit the flooding of packets to a small region.

### 3.3 Zone Routing

A routing zone is defined for each node separately, and the zones of neighboring nodes overlap. A k-hop routing zone of node S can be defined as a connected topological subgraph, on which node S is aware of the route to any other node [5]. The k-hop zone thus includes the nodes, whose distance (not physical) from the node in question is at most k-hops, therefore it can be of any shape. The nodes of a zone are divided into border nodes and interior nodes. Border nodes are nodes which are exactly k hops away from the node in question. The nodes which are less than k hops away are interior nodes. In fig. 2, the nodes G, D and M are border nodes and rest all are interior nodes and the node N, 4 hops away from S, is outside the routing zone. However node L is within the zone, since the shortest path up to L with length 3 is less than the maximum routing zone hops.

In an ad-hoc network, it can be assumed that most of the traffic is directed to nearby nodes. Therefore, the proactive scope is reduced to a small zone around each node in the SEELAMP protocol. In a limited zone, all nodes proactively keep track of their neighbor nodes within the zone, hence in a zone routing network, each node maintains a proactive unicast route to every other node within the specified zone. In the proposed protocol the routing is initially established with proactively prospected routes within the zone and then outside the zone, serves the demand from selectively activated nodes through reactive approach using the physical location of the nodes. Therefore, route requests can be more efficiently performed without exploiting the flooding in the network.

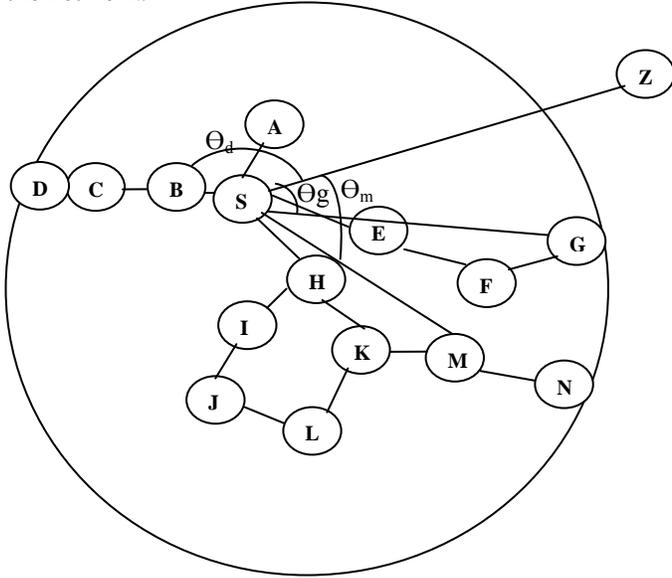

Fig.2. 3 hop Routing zone at node S

As the zone radius is significantly smaller than the network radius, the cost of learning the zones' topologies is a very small fraction of the cost required by a global proactive mechanism. Zone routing is also much cheaper (in terms of control traffic and congestion) and faster than a global reactive route discovery mechanism, as the number of nodes queried in the process is very small [20]. A bigger proactive zone can be selected for comparatively stable topology where the updates of topology are done on topology change only otherwise a smaller zone can be preferred.

### 3.4 Data Structure

SLS represents a fully distributed location service, i.e., each node in the network maintains a part of the overall location database. In order to facilitate the location service, each node has some data structures in addition to those needed for the routing algorithm. To reduce the redundancy in the data structure we have modified the data structure used by SLS, which are as follows:

First, a localized "zone neighbor table" is maintained by each node, which contains routing entries for each neighbor within the k-hop routing zone as shown in table 1. Each routing entry contains the IP of neighbor nodes, their location, next immediate hop, hop count to this particular node and a timestamp indicating when the entry was added or updated. In case of enough space available in a node, it may store the entries of other nodes about which it learnt by passively listening on the network in addition to its zone neighbors.

In order to exchange location information on the network, SLS uses four special packet types which are exchanged in the same way as data packets.

A LOCN packet as shown in fig. 5 is broadcasted by a node in its zone when it wants to inform other node(s) of its location. It contains the IP, location (latitude and longitude) of the source node and a timestamp.

| Source Node | | | Timestamp |
|---|---|---|---|
| | | | TS |
| IP | Latitude | Longitude | |
| SRC_IP | SRC_LOC | SRC_LNGT | |

Figure 5. Format of LOCN packet

When a node receives a LOCN packet from another node it unicasts back a location acknowledgement packet LACK as shown in fig. 6. This packet contains the IP and location of the source node, the IP and location of the node acknowledging receipt of a LOCN and a timestamp.

Obtaining the locations of the mobile nodes, distance d between two mobile nodes can be calculated using (1) and slope theta is calculated using (2).

$$d = \sqrt{(x2-x1)^2 + (y2-y1)^2} \quad (1)$$

$$\theta = \tan^{-1}\frac{(y2-y1)}{(x2-x1)} \quad (2)$$

where (x1,y1) and (x2,y2) are the locations of two mobile nodes.

Table 1: Zone Neighbor Table maintained by each node

| Neighbor Node IP | Location | | Immediate Next Hop NXT_HOP | Hop Count HOP_CNT | Timestamp TS |
|---|---|---|---|---|---|
| | Latitude LAT | Longitude LNGT | | | |
| 222.24.15.06 | 420 10' E | 560 40' S | 222.24.15.15 | 3 | 15:09 PM |
| 222.24.15.11 | 550 10' W | 340 33' S | 222.24.15.31 | 2 | 15:15 PM |
| 222.24.15.20 | 230 26' E | 150 14' N | 222.24.15.19 | 3 | 15:24 PM |
| 222.24.15.29 | 450 30' N | 430 20' E | 222.24.15.43 | 1 | 15:42 PM |

| Ack. Node | | | Source Node | | | Time-stamp |
|---|---|---|---|---|---|---|
| IP | Latitude | Longitude | IP | Latitude | Longitude | TS |
| ACK_IP | ACK_LAT | ACK_LNGT | SRC_IP | SRC_LAT | SRC_LNGT | |

Figure 6. Format of LACK packet

| Request Node | | | Multicast Group IP | Join Flag | Time-Stamp |
|---|---|---|---|---|---|
| IP | Latitude | Longitude | MG_IP | JF | TS |
| RQ_IP | RQ_LAT | RQ_LNGT | | | |

Figure 7. Format of MGREQ packet

| Multicast Group Tree Member | | | Request node | | | Time stamp |
|---|---|---|---|---|---|---|
| IP | Latitude | Longitude | RQ_IP | Latitude | Longitude | TS |
| TM_IP | TM_LAT | TM_LNGT | | RQ_LAT | RQ_LNGT | |

Figure 8. Format of MGRPL packet

When a node wants to search for an existing multicast group it broadcasts a multicast group request packet MGREQ, shown in fig. 7, within its zone. This packet contains the IP and location of the request node, IP of the multicast group and a timestamp.

A location reply packet MGRPL as shown in fig. 8 is sent in response to a MGREQ packet. The MGRPL packet contains the IP and location of the multicast group tree member, the IP and location of the request node, and a timestamp.

Entries are added and updated in ZNT on the reception of LOCN, LACK, MGREQ and MGRPL. Besides Zone Neighbor Table (ZNT), for the purpose of routing information each node maintains Multicast Tree Table (MTT) as shown in table 2 and Request Table (RT) as shown in table 3.

Each entry of Multicast Tree Table contains the multicast group IP address, multicast group leader IP address, hop count to multicast group leader, next hops and timestamp. This table has entries for all those multicast groups of which group the node is a member. The Next Hops field is a linked list of structures, each of which contains the following fields:
- Next Hop IP Address
- Link Direction
- Activated Flag

Table 2: Multicast Tree Table

| IP Multicast group MG_IP | IP Multicast Group Leader MGL_IP | Hop Count HOP_CNT | Immediate Next hop NXT_HOP | Time-stamp TS |
|---|---|---|---|---|
| 224.30.15.10 | 222.24.15.50 | 8 | 222.24.15.05 | 15:39 PM |
| 224.30.10.10 | 222.24.15.65 | 9 | 222.24.15.11 | 15:45 PM |
| 224.30.10.15 | 222.24.15.36 | 7 | 222.24.15.10 | 15:04 PM |
| 224.30.10.50 | 222.24.15.45 | 10 | 222.24.15.20 | 15:12 PM |

Table 3: Request Table

| IP Multicast group MG_IP | IP Multicast group Tree Member TM_IP | Location of Multicast group Tree Member | | Time-stamp TS |
|---|---|---|---|---|
| | | Latitude TM_LAT | longitude TM_LNGT | |
| 224.30.15.10 | 222.24.15.06 | 420 10' E | 560 40' S | 13:41 PM |
| 224.30.10.10 | 222.24.15.11 | 550 10' W | 340 33' S | 14:38 PM |
| 224.30.10.15 | 222.24.15.20 | 230 26' E | 150 14' N | 13:24 PM |
| 224.30.10.50 | 222.24.15.29 | 450 30' N | 430 20' E | 14:12 PM |

The direction of the link is relative to the location of the group leader. UPSTREAM is a next hop towards the group leader, and DOWNSTREAM is a next hop away from the group leader [5]. An entry is added to the table when the node becomes a multicast group member.

A request table is maintained by all those nodes that support multicast. An entry in this table contains multicast group IP address, requesting node (actually a tree member node) IP address, requesting node location and a timestamp. On reception of MGREQ with join flag set (MGREQ-J) from a request node an entry is made in this table [10].

### 3.5 Neighborhood Connectivity Updation

Nodes learn of their neighbors through transmission of LOCN, LACK, MGREQ and MGRPL packets used by SLS. In SEELAMP, instead of beaconing, a node broadcasts LOCN packet with TTL value equal to k hops whenever it enters into a network or whenever it moves significantly from the previous location, to inform other node(s) of its location and these neighbors unicasts the LACK packet to the sending node to get update their locations. When a node sends an LACK packet to some node, all of its neighbors hear the transmission and maintains this node as their neighbor in the ZNT with the appropriate value of hop count. A node also learns of its neighbours by promiscuous snooping on the channel for detecting activities of neighbors. A soft state approach is followed to remove stale routes from the ZNT. Entries are removed after a certain interval of time with respect to time stamp in order to clear a node's table of possibly outdated information.

### 3.6 Multicast Tree Creation and Maintenance

SEELAMP maintains a bi-directional shared multicast tree for each multicast group, consisting of the members of the multicast group and several routers. Each multicast group has a unique multicast group IP (address) [8] and a group leader. The group member that first constructs the tree is designated as the group leader or the primary root of the tree [10].

**3.6.1 Joining the existing tree -** A request node, that wants to join the multicast group, will first look for the existing tree of the multicast group. The node broadcasts a MGREQ message with join flag set to the nodes within its zone in search of multicast group IP. All nodes of the zone search the multicast group IP in their multicast tree table first and then in request table and a node having a matched entry replies MGRPL back unicastly. The MGRPL is sent using the reverse route maintained during the traversing of MGREQ packet. In case of no entry matches in the multicast tree table or request table of all the neighbor nodes, the request node searches the tree existence outside the zone.

**3.6.2 Link grafting -** After receiving the MGRPL following the forward route the request node sends the GRAFT message to confirm the join process to the node from which it received the MGRPL message. The GRAFT message will activate the tree link between the request node and the node which sent the MGRPL message and the request node becomes the tree member. Request node also updates its request table and multicast tree table.

**3.6.3 Creating a new tree for a new multicast group -** Once the whole network is traced in search of multicast group and still no MGRPL is received by the request node, it assumes that the requested multicast group does not exist. It then declares itself the leader of the multicast group and becomes the primary root of the tree and broadcasts this information to all nodes within its zone.

**3.6.4 Maintenance of the tree -** The robustness of the multicast tree is adversely affected with the time as individual links are repaired only when broken. Over a period of time due to high mobility among the nodes the overall structure of the tree would be far from optimal, hence making the tree susceptible to even more link breakages. In SEELAMP, the tree is updated regularly and also the preventive maintenance is done which kept the tree robust.

**3.6.4.1. Tree Updation -** In order to maintain the tree structure even when nodes move, group members periodically send tree_update requests to the backup root node to lessen the load on the primary root node. The multicast tree can be updated using the path information included in the tree_update request messages. If any change is found in the path the back up root node sends an update message to the primary root node to notify about the change so that the changes in the topology also reflect in the tree structure. Tree_update need to be initiated by leaf nodes only as each uplink next hop puts its own uplink on the tree update message, therefore contains all uplinks as it travels towards backup root node. The period must be carefully chosen to balance the overhead associated with tree update and the delay caused by the tree not being timely updated when nodes move [4], [8], [9].

**3.6.4.2 Preventive Multicast tree Maintenance -** A preventive approach is being proposed for tree reconstruction prior to link breakages using the following methods:

**a. Leaving the tree-** In the proposed protocol a non-leaf node wishing to move out of the multicast tree, will broadcast an alarm message to all of its neighbors with TTL value 1 before sending the Leave message. It then compares the location of nodes in its ZNT and passes all of its routing information to a nearest node which is not a tree member. Thus new links are grafted on the tree from the upstream node and downstream nodes of the leaving node to the newly found neighbor node. The downstream node sends tree_update to the backup root node. All the future transmissions follow the path with newly discovered link. In case of leaf node or a normal network node, the node simply sends the leave message to its one hop neighbor nodes. All the neighbor nodes receiving the alarm packet from any node also remove the related entry from their ZNT and also from request table, if the entry with IP of leaving node exists there. In case of primary root mobility, the primary root sends the alarm message to back up root notifying it to take the control of the tree and passes its all routing information to the back up root. Upon receiving the alarm message, the back up root updates its downstream next hops to the downstream next hops of the primary root node. It also selects a new back up root for its replacement after it resumes as primary root node.

**b. Power Resource Depletion** - Another proactive measure can be taken in case of the complete depletion of the power sources of a node of the multicast tree. Route is reconfigured quickly in case of a node goes off because of complete drainage of its energy sources. The power sources of the nodes in the multicast tree are examined periodically (frequency of examination is doubled in case of primary root node) and if the power source of a node goes below a threshold value, a new link is discovered prior to its failure, and the links to this node are deleted from the multicast tree. New link is searched in the same way as in case of leaving the tree process.

The latency in finding new route in case of nodes failure is reduced by reconfiguring the routes using preventive approach before the failure of the node.

**3.6.4.3 Multicast Tree Repair** - When a link breakage is detected, the downstream node of the break (node farther away from the group leader) initiates to repair the link by broadcasting a MGREQ-J within the zone. Only a tree node with lesser hop count to the leader (that is nearer to the group leader) may respond to this MGREQ. If the node receives a reply it then grafts a new branch using GRAFT message up to the node which sent the MGRPL.

## 3.7 Data Transmission Process

**3.7.1 Tree member search within zone** - Proactive topological routing operates within the k-hop routing zone. When a node wants to route a data packet or join a multicast group, it first checks the multicast tree member in the zone by broadcasting a MGREQ packet with multicast group IP within its k-hop routing zone (TTL=k). All nodes of the zone search the multicast group IP in their multicast tree table. If an entry of the multicast tree table matches, then the node unicasts MGRPL to the request node with its own IP, latitude and longitude in the multicast tree member IP, latitude and longitude fields of the MGRPL. After receiving the MGRPL the request node broadcasts a stop search signal to all nodes in its zone. The request node then sent the packet along the forward route created during MGRPL transmission.

**3.7.2 Tree member search outside zone** - In case of no entry found in the multicast tree table of the zonal nodes, these nodes search the multicast IP in the request table. If any entry of the request table matches, then the node unicasts MGRPL to the request node by putting the IP, latitude and longitude of the matched entry node in the multicast tree member IP, latitude and longitude fields of the MGRPL. The matched entry node, in the request table, is actually the tree member node outside the zone.

In case of no entry matches in the multicast tree table and the request table of the nodes in the zone and the network diameter is still not reached, then the request node sends a small signal to its border nodes (in case of no border nodes, nodes with k-1 hop as shown in fig.2) to forward the cached copy of the MGREQ to all their border nodes for further searching of the multicast group in the whole network as shown in fig.1. The border nodes then broadcast the MGREQ packet to all the nodes in their zone. These nodes further search the multicast group IP in their multicast tree table and request table and a node having a matched entry in either table replies MGRPL back unicastly to the request node.

**3.7.3 Diffused reactive routing** - If the tree member node is found through multicast tree table of any node outside the zone then the request node sent the packet along the forward route to this tree member. If the tree member node is found through request table of any node within or outside the zone then a route upto the tree member is find out by directional diffused reactive routing using its location information. For that the request node selects only those nodes lying on the perimeter of its k-hop zone which are in the direction of the tree member, hence geographically closer to the tree member. As shown in fig. 2, node S selects the border nodes G and M only as their slope magnitude with the direction of tree member θg and θm are less than the threshold slope magnitude θt while the slope magnitude of D i.e. θd is more than θt. After selecting the nodes, the packet is forwarded to the next hops towards the border nodes. In case of sparse network, if there is no border node in the zone then the route search packet is forwarded to only those farthest neighbor nodes in the zone which are having slope less than the threshold slope with the direction of the tree member.

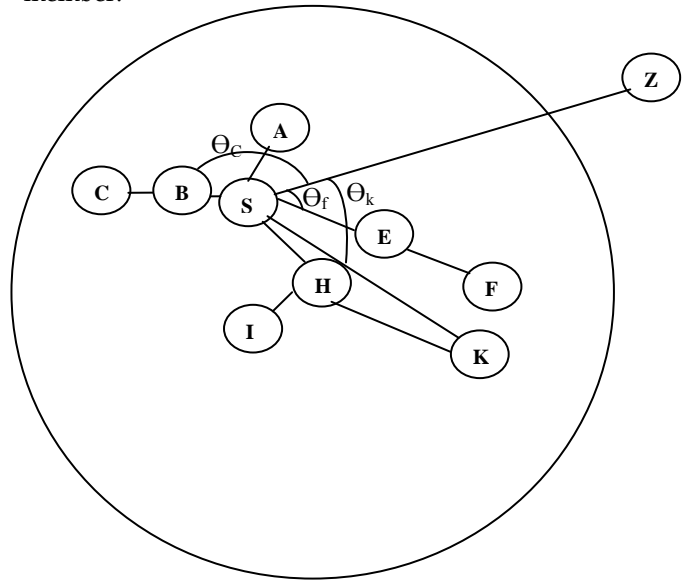

Fig.3. 3 hop Routing zone at node S with no border nodes

As shown in fig. 3, node S forwards the packet towards the neighbor nodes F and K through next hops E and H only as their slope magnitude θf and θk are less than the threshold slope magnitude θt while the slope magnitude of C i.e. θc is more than θt. If no such nodes are found then the route search packet is forwarded to all neighbor nodes within the zone which further forwards the packet to their border nodes in the direction of the tree member as described above. As shown in fig. 4, node S forwards the packet towards all the neighbor nodes C, F and H through next hops B, E and H as no node is having the slope less than the threshold slope magnitude θt. Thereafter these border or farthest nodes will forward the route search packet to the border nodes of their respective k-hop zones in the direction of tree member only. This process goes on until the packet reaches to the tree member specified in the MGRPL packet. After accepting the first copy of the packet rest copies are discarded by the tree member. This tree member now replies back the MGRPL to the request node. The request node transmits the data packet to the tree member along the forward route.

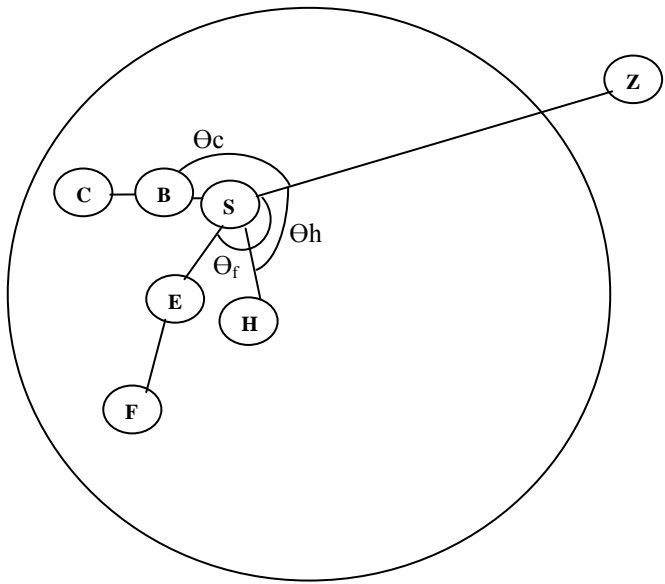

Fig.4. 3 hop Routing zone at node S with no slope condition satisfying node

Since the traffic would be forwarded only through limited nodes to tree members for route discovery using diffused reactive routing it effectively reduces the traffic and saves the bandwidth a lot.

SEELAMP algorithm searches the multicast group tree member in the zone of the intermediate node. It searches the possibility of tree member by searching the multicast group IP in the multicast tree table and request table of each neighbor node in the zone. In case of no match found with in the zone it repeats the search outside the zone.

Algorithm: SEELAMP (MGREQ, ZNT_Size, MTT_Size, RT_Size)
// RQ is the request node, ZNT is the zone neighbor table having entries for all neighbors in its zone and some other neighbors in case of free room, MTT is the multicast tree table maintained by each node having entries of those multicast groups to whom the node belongs as a tree member. RT is the request table having entries of all those nodes which made request for any multicast group. ZNT_Size, MTT_Size and RT_Size are the maximum entries in the corresponding tables and MGREQ is the multicast group request packet.

```
TTL=0;
repeat {
  if ( ZNT_Size = = 1 ){
// forward the data packet to only entry in the zone neighbor table
     forward ( MGREQ, ZNT[1].IP);
     TTL+=1;
  }
  else {
//search the availability of multicast group member in the neighborhood by checking the Multicast Tree Table
    repeat {
    i = 1; j=1; m=1; n=0;

    do {
     if (ZNT[i].MTT[j].MG_IP = = MGREQ.MG_IP )
     {
        n=1;
        exit();
     }
     j++;
    } while (j < MTT_Size)

    if (n= =1)
    {
      MGRPL.TM_IP = ZNT[i].IP;
      MGRPL.TM_LAT = ZNT[i].LAT;
      MGRPL.TM_LNGT = ZNT[i].LNGT;
      MGRPL.RQ_IP = MGREQ.RQ_IP;
      MGRPL.RQ_LAT = MGREQ.RQ_LAT;
      MGRPL.RQ_LNGT = MGREQ.RQ_LNGT;
      forward ( MGRPL, MGREQ.RQ_IP);
      if (MGREQ.JF= =1)
         forward(GRAFT, ZNT[i].IP);
      TTL+= ZNT[i].HOP_CNT;
      exit();
    }
//search the availability of multicast group member outside the neighborhood by checking the Route Table

    do {
     if (ZNT[i].RT[m].MG_IP = = MGREQ.MG_IP )
     {
       n=2;
       exit();
     }
```

```
        m++;
    } while (m < RT_Size)

    if (n= =2) {
//send MGRPL by putting the tree member 's IP and its
location into the tree member 's IP and location field of
the MGRPL
        MGRPL.TM_IP = RT[m].TM_IP;
        MGRPL.TM_LAT = RT[m].TM_LAT;
        MGRPL.TM_LNGT = RT[m].TM_LNGT;
        MGRPL.RQ_IP = MGREQ.RQ_IP;
        MGRPL.RQ_LAT = MGREQ.RQ_LAT;
        MGRPL.RQ_LNGT = MGREQ.RQ_LNGT;
        forward ( MGRPL, MGREQ.RQ_IP);

        TTL+= ZNT[i].HOP_CNT;
//find the route up to the tree member using directional
diffused forwarding routing in the direction of tree
member
        repeat {
            directional diffused forwarding algorithm (
MGREQ, TTL)
        } until (tree member found)
        if (tree member found) {
            MGRPL.TM_IP = TM_IP;
            MGRPL.TM_LAT = TM_LAT;
            MGRPL.TM_LNGT = TM_LNGT;
            MGRPL.RQ_IP = MGREQ.RQ_IP;
            MGRPL.RQ_LAT = MGREQ.RQ_LAT;
            MGRPL.RQ_LNGT = MGREQ.RQ_LNGT;
            forward ( MGRPL, MGREQ.RQ_IP);
            if (MGREQ.JF= =1)
                forward(GRAFT, TM_IP);
            exit();
        }
    } //(n= = 2)
    i+ +;
} until ((i > ZNT_Size) or (n= =1) or (n= = 2))
} //else

// tree member node is not found within the MTT or RT
of neighbor nodes, so forward MGREQ to all border
nodes through next hop
if (n= =0) {
    i=1;
    repeat {
        if ( ZNT[i].HOP_CNT= =k) {
            forward ( MGREQ, ZNT[i].IP);
        }
    TTL+=k;
    i++;
    } until (i > ZNT_Size)
} //(n= = 0)

} until (TTL >Net_Dia)

//request node declares itself the group leader if the
multicast group does not exist in the network
if (TTL = =Net_Dia or MGREQ.JF = =1){
    MTT[MTT_Size+1].MG_IP = MGREQ.MG_IP;
    MTT[MTT_Size+1].MGL_IP = MGREQ.RQ_IP;
    MTT[MTT_Size+1].HOP_CNT = 0;
    RT[RT_Size+1].MG_IP = MGREQ.MG_IP;
    RT[RT_Size+1].MGTM_IP = MGREQ.RQ_IP;
    RT[RT_Size+1].MGTM_LAT= MGREQ.RQ_LAT;
    RT[RT_Size+1].MGTM_LNGT=
                            MGREQ.RQLNGT;
}
```

Directional diffused forwarding algorithm forwards the MGREQ to the border nodes or the farthest nodes in the zone in the direction of the tree member. It selects only those border nodes whose slope with the direction of tree member is less than the threshold value of the slope. Slope can be find out by (2) using the latitude and longitude of the nodes.

```
Algorithm: Directional diffused forwarding algorithm (
MGREQ, TTL)
// forwards the MGREQ to the border nodes in the
direction of the tree member
i=1; n=0;
// border nodes with slope less than the threshold with
the direction of the tree member are selected by the
algorithm
repeat
    repeat
        if ( (ZNT[i].HOP_CNT = =k)  &&
        (slope(ZNT[i])< thresh_slope)) {
            forward ( MGREQ, ZNT[i].IP);
    n=1;
        }
        i++;
    until (i= =ZNT_Size)
    if (n = =  1){
        TTL+=k;
    else {
        --k;
        i=1;
    }
until (k= = 0 or n= =1)
return (TTL);
```

## 4 PERFORMANCE COMPARISON

| Protocols / Characteristics | Mesh-based Multicast | Per-source tree Multicast | Shared Tree multicast | SEELAMP |
|---|---|---|---|---|
| Packet Delivery Ratio | High due to redundant paths | Low due to link errors | Low due to link errors | Moderate due to preventive maintenance and local link repair |
| Network Structure Construction & Maintenance overhead | High due to forwarding mesh | High due to one tree per source per group | Low due to only one shared tree per group | Even less than Shared tree multicast due to only one shared tree per group and backup root node and preventive maintenance |
| Latency due to link error | Low due to redundant links | High due to new tree construction | High due to new tree construction | Moderate due to preventive maintenance and local link repair |
| Energy Consumption | High due to broadcast flooding | Low due to unicast transmission | Low due to unicast transmission | Low due to location aware and diffused reactive routing for finding the route |
| Load Distribution | Good due to distribution of load to forwarding nodes | Good due to load distribution to multiple trees | Poor due to single shared multicast tree | Better than Shared tree multicast due to backup root node |

## 5 CONCLUSION

The Scalable Energy Efficient Location Aware Multicast Protocol is compared with Mesh Based Multicast, Per-Source Tree Multicast and Shared Tree Multicast protocols which are based on two popular multicast routing i.e. mesh-based and tree-based methodology for ad hoc networks. Comparison was made on various parameters like Packet Delivery Ratio, Network Structure Construction & Maintenance overhead, Latency due to link error, Energy Consumption and Load Distribution.

SEELAMP eliminates the drawbacks of mesh based protocols, per-source tree and even that of shared tree protocols. It reduces the latency problem due to directional diffused forwarding routing and also the network partition problem when a link error occurs due to the failure of primary root. It also avoids any coupling with an underlying unicast routing protocols without incurring any extra overhead due to the incorporation of SLS for obtaining the physical location of the nodes and for other information sharing.

Backup root also facilitates reduction in overhead in case of SEELAMP otherwise required for tree reconstruction and tree maintenance. This result in improved packet delivery ratio and energy balance compared to the conventional shared tree multicast (STM) due to preventive maintenance, local link repair and also because it can switch to the backup root when the primary root is overloaded or becomes invalid.

Scalability is achieved due to the shared tree multicast routing protocol as single tree maintenance of all group members is easier than the maintenance of number of trees in case of source based multicast routing protocol.

## ACKNOWLEDGMENT

I am grateful to Dr. Ranjit Biswas, Professor, and Head of Department of Computer Science, Institute of Technology and Management, Gurgaon, India for useful guidance and motivation. I am also thankful to my family for kind support and affection.

## REFERENCES

[1] M. Gerla, C.-C. Chiang, and L. Zhang, "Tree Multicast Strategies in Mobile, Multihop Wireless Networks," Baltzer/ACM Journal of Mobile Networks and Applications (MONET), Vol. 3, No. 3, pp. 193-207, 1999.

[2] J.J Garcia-Luna-Aceves and E.L. Madruga, "The Core-Assisted Mesh Protocol," IEEE Journal on Selected Areas in Communication, vol. 17, no. 8, August 1999.

[3] A.K. Sharma and Amit Goel, "Moment to Moment Node Transition Awareness Protocol (MOMENTAP)", International Journal of Computer Applications (IJCA) Special Issue, IASTED, Vol. 27/1, Jan 2005, pp. 1-9.

[4] Elizabeth M. Royer and Charles E. Perkins, "Multicast Operation of the Ad-hoc On-Demand Distance Vector Routing Protocol", in Proceedings of the 5th Annual ACM/IEEE International Conference on Mobile Computing and Networking (Mobicom'99), Seattle, WA, USA, August 1999, pages 207-218.

[5] Hui Cheng and Jiannong Cao, "A Design Framework and Taxonomy For Hybrid Routing Protocols in Mobile Ad Hoc Networks", IEEE Communications, Surveys 3rd Quarter 2008, Volume 10, No. 3.

[6] Pariza Kamboj, A.K.Sharma, "Location Aware Reduced Diffusion Hybrid Routing Algorithm (LARDHR)", accepted for ICETET 09, Nagpur, India.

[7] Stephen Mueller, Rose P. Tsang and Dipak Ghosal, " Multipath Routing in Mobile Ad Hoc Networks: Issues and Challenges".

[8] Yufang Zhu and Thomas Kunz, "MAODV Implementation for NS – 2.26" Communications and Networking in China, 2006. ChinaCom apos;06. First International Conference on Volume, Issue 25-27, pp:1 - 5

[9] M. Liu, R. R. Talpade, A. McAuley, and E. Bommaiah, "AMRoute: Adhoc Multicast Routing Protocol," Technical Report, vol. TR 99-8, The Institute for Systems Research, Univesity of Maryland, 1999.

[10] Pariza Kamboj, A.K.Sharma, "MAODV-PR: A Modified Mobile Ad Hoc distance Vector Routing Protocol with Proactive Route Maintenance", VOYAGER - The Journal of Computer Science & Information Technology, Vol. 6, No. 1, Jan-June 2008, pp. 35-41.

[11] J. Li, J. Jannotti, D. S. J. D. Couto, D. R. Karger, and R. Morris, "A scalable location service for geographic ad hoc routing," presented at the ACM/IEEE MobiCom, Boston, MA, Aug. 2000.

[12] N. K. Guba and T. Camp, Recent work on GLS: a location service for an ad hoc network, Proceedings of the Grace Hopper Celebration (GHC), 2002.

[13] S. Zahoor Ul Huq1, Dr. K.E. Sreenivasa Murthy, D. Kavitha, B. Satyanarayana, "EMMR: A Multicast Protocol for Mobile Ad Hoc Networks", IJCSNS International Journal of Computer Science and Network Security, VOL.9 No.4, April 2009

[14] Sangman Moh, Chansu Yu, Ben Lee, and Hee Yong Youn, "Energy Efficient and Robust Multicast Protocol for Mobile Ad Hoc Networks", Proceedings of the 2002 Pacific Rim international Symposium on Dependable Computing (December 16 - 18, 2002). Proceedings of IEEE Computer Society, Washington, DC, 145.

[15] S. Basagni, I. Chlamtac, V. R. Syrotiuk, and B. A. Woodward, "A distance routing effect algorithm for mobility (DREAM)," presented at the ACM/IEEE MobiCom'98, Oct. 1998.

[16] Y. B. Ko and N. H. Vaida, "Location-aided routing (LAR) in mobile ad hoc networks", presented at the ACM/IEEE MobiCom'98, Oct. 1998.

[17] Xiaojing Xiang, Xin Wang, Zehua Zhou, "An Efficient Geographic Multicast protocol for Mobile Ad Hoc Networks", In Proceedings of the 2006 international Symposium on on World of Wireless, Mobile and Multimedia Networks (June 26 - 29, 2006). International Workshop on Wireless Mobile Multimedia. IEEE Computer Society, Washington, DC, 73-82. DOI= http://dx.doi.org/10.1109/WOWMOM.2006.22

[18] Song Guo, Member, IEEE, and Oliver Yang, Senior Member, IEEE, "Maximizing Multicast Communication Lifetime in Wireless Mobile Ad Hoc Networks", IEEE Transactions on Vehicular Technology, vol. 57, no. 4, July 2008

[19] L. Ji and M. S. Corson. A Lightweight Adaptive Multicast Algorithm. Proceedings of IEEE GLOBECOM, pages 1036-1042, Sydney, Australia, December 1998.

[20] Aniruddha Rangnekar, Ying Zhang,Ali A. Selcuk, Ali Bicak, Vijay Devarapalli, Deepinder Sidhu, "A Zone-Based Shared-Tree Multicast Protocol for Mobile Ad Hoc Networks", In Vehicular Technology Conference, 2003, 2003.

[21] S.-J. Lee, M. Gerla, and C.-C. Chiang, "On-Demand Multicast Routing Protocol," in Proceedings of IEEE WCNC'99, September 1999.

[22] J. E. Wieselthier, G. D. Nguyen, and A. Ephremides, "Algorithms for Energy-Efficient Multicasting in Ad Hoc Wireless Networks," Proc. of Military Communication Conference (MILCOM 1999), Vol. 2, pp. 1414-1418, Nov. 1999.

## About the Authors

**Pariza Kamboj**
Ms. Pariza Kamboj received her M.Tech (Comp. Sc. & Engg.) with Hons. from Kururkshetra University (KU), Kurukshetra, Haryana (India) in the year 2006. From July 1996 to Mar 2007, she served at JMIT, Radaur, Yamuna Nagar, Haryana (India). From Apr 2007 to Nov 2007, she served as Assistant Professor at CITM, Faridabad, Haryana (India). Since Dec 2007, she is working as Assistant Professor at ITM, Gurgaon (India) Computer Engg. & Information Technology. Currently she is pursuing her Ph.D. in the area of Mobile Ad Hoc Networks.

**Ashok Sharma**
Prof. A.K.Sharma received his M.Tech (Comp. Sc. & Tech) with Hons. from University of Roorkee (India) in the year 1989 and Ph.D. (Fuzzy Expert System) from JMI, New Delhi (India) in the year of 2000. From July 1992 to April 2002, he served as Assistant Professor and became Professor in Computer Engg at YMCA Institute of Engineering, Faridabad (India) in April 2002. He received his 2nd Ph.D. in Information Technology from Indian Institute of Information Technology & Management, Gwalior (India) in the year 2004. His research interest includes Fuzzy System, Knowledge Representation, and Computer Networks & Internet Technology.